\documentclass[aps,pre,twocolumn,showpacs]{revtex4}
\usepackage{graphicx}

\begin{document}

\title{Error-correcting codes on scale-free networks}

\author{Jung-Hoon Kim and Young-Jo Ko}
\affiliation{Future Technology Research Division,
         Electronics and Telecommunications Research Institute,
         Daejeon 305-350, Korea}
\date{\today}

\begin{abstract}
We investigate the potential of scale-free networks as
error-correcting codes. We find that irregular low-density
parity-check codes with highest performance known to date have
degree distributions well fitted by a power-law function $p(k)\sim
k^{-\gamma}$ with $\gamma$ close to 2, which suggests that codes
built on scale-free networks with appropriate power exponents can
be good error-correcting codes, with performance possibly
approaching the Shannon limit. We demonstrate for an erasure
channel that codes with power-law degree distribution of the form
$p(k)=C(k+\alpha )^{-\gamma}$, with $k \geq 2$ and suitable
selection of the parameters $\alpha$ and $\gamma$, indeed have
very good error-correction capabilities.
\end{abstract}


\maketitle

A variety of complex networks \cite{Alb1} exhibit a topological
structure in which the connectivity between their constituent
nodes follows a simple power law. Examples of such scale-free
networks include the Internet \cite{Fal,Cal}, the World Wide Web
(WWW) \cite{Alb2,Hub}, social networks \cite{New1}, metabolic
networks \cite{Jeong}, etc. Extensive studies have been made to
understand the topological features and evolving dynamics
\cite{Bara,Krap,Doro} of these networks. While many intriguing
properties concerning the structural aspect of complex networks
have been revealed thanks to these efforts, relatively little has
been known about the effects of specific connectivity structures
on networks' functional behavior \cite{Stro}. In order to properly
operate under a certain environment or, in a more active sense, to
successfully accomplish a given task, a complex network may favor
one particular structure over another. For practical applications,
it now appears that more attention need be paid to the {\em
functional aspect} of these complex networks viewed as whole
systems or organisms working for particular purposes.

Recent advances in channel coding theory have led to the
perception that the state-of-the-art capacity-approaching codes,
such as Turbo codes \cite{Berr} and low-density parity-check
(LDPC) codes \cite{Gall,Mac1,Rich}, can be understood in terms of
graphs (or networks) consisting of nodes and edges \cite{Forney}.
The function of these graphs is to carry out error correction,
i.e. to recover original data transmitted over noisy channels, by
iteratively passing certain messages through edges connecting
neighboring nodes. The art of developing a high-performance
error-correcting code lies in designing a connectivity structure
of a graph in such a way as to make the code built on it perform a
desired function. One very important issue concerns finding the
connectivity distribution that achieves the Shannon's capacity
limit. Most attempts, however, have been limited to numerical
optimization, and a complete understanding of the connectivity
structure specific to capacity-achieving codes is still lacking.
Inspired by the ubiquitous nature of scale-free networks, one may
ask whether the connectivity structure of scale-free networks
could offer any insight into seeking good graph-based codes.

In this paper we address the question whether scale-free networks
whose connectivity distribution follows a power law can function
effectively as good error-correcting codes. The codes built on
scale-free networks considered here are basically LDPC codes in
that the associated parity-check matrices are sparse and that the
belief propagation algorithm \cite{Gall,Mac1,Rich} is employed for
decoding. We first show that the degree distributions of LDPC
codes with highest performance known to date are well fitted by
power-law functions. Motivated by this finding, we generate a
degree distribution according to the function $p(k)=C(k+\alpha
)^{-\gamma}$ and fine tune the parameters $\alpha$ and $\gamma$ to
maximize the code's performance. We investigate error-correction
capability of these codes over a binary erasure channel and
compare them with the Tornado code \cite{Luby}, the first
commercialized LDPC code.

An LDPC code can be represented by a bipartite graph in which
there are two different types of nodes: variable nodes and check
nodes. Nodes of one type are connected by edges only to nodes of
the other type. Variable nodes are associated with data bits, and
check nodes examine whether the variable nodes connected to them
satisfy parity-check equations. Error correction of corrupted data
bits is performed by passing certain messages, e.g. likelihood
ratios, through edges back and forth between variable and check
nodes. It is known from the density evolution analysis \cite{Rich}
that, under the assumption of a tree-structured random graph with
no closed loops, the error-correction capability of a code is
solely determined by the degree distribution.

\begin{figure}
\includegraphics[width=0.95 \columnwidth]{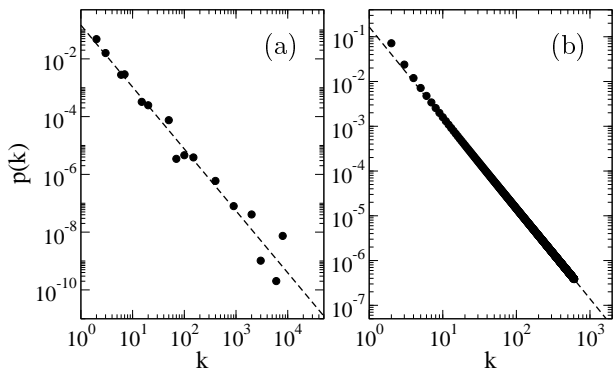}
\caption{\label{fig1} Degree distributions for high-performance
LDPC codes. (a) The LDPC code designed by Chung {\it et al.}
(Table \ref{tab2} of \cite{Chung}). (b) The Tornado code with
maximum degree $d_{max}=610$ \cite{Luby,Shok}. The best fitting
lines have slopes (a) $\gamma \simeq 2.14$ and (b) $\gamma \simeq
2.02$. }
\end{figure}

We begin by inspecting the degree distributions of some
high-performance LDPC codes. Figure \ref{fig1}(a) shows the
variable-node degree distribution of the LDPC code designed by
Chung {\it et al.}~\cite{Chung}, which has been optimized for an
additive white Gaussian noise channel and approaches the Shannon
limit within 0.0045 dB, presently the world record. Here, in order
to obtain a meaningful distribution from the irregularly spaced
data $\lambda (k_i)$ in Table \ref{tab2} of Chung {\it et
al.}~\cite{Chung}, we took a local average over a bin of length
$(k_{i+1} - k_{i-1})/2$:
\begin{equation}\label{eq1}
p(k_i) = \frac{P(k_i)}{(k_{i+1} - k_{i-1})/2},
\end{equation}
where $P(k_i)$ is the fraction of nodes with degree $k_i$ and is
given by $P(k_i) = C\lambda (k_i)/k_i$, in which $\lambda (k_i)$
is the fraction of edges connected to a variable node of degree
$k_i$ and $C$ is a normalization constant. It can be seen from
Fig.~\ref{fig1}(a) that the degree distribution is well fitted by
a power-law function $p(k) \sim k^{-\gamma}$ with $\gamma \simeq
2.14$. A more dramatic correspondence is observed for the
variable-node degree distribution of the Tornado code
\cite{Luby,Shok}, as apparently seen in Fig.~\ref{fig1}(b). The
Tornado code has been optimized for an erasure channel and has a
Poisson distribution for its check-node degree distribution. The
power-law function that best fits the variable-node degree
distribution is found to have an exponent $\gamma \simeq 2.02$,
and the fitting appears to be nearly perfect for large $k$. We
also find that the right-regular sequence of Shokrollahi
\cite{Shok} that slightly beats the Tornado sequence is well
fitted by a similar power-law function.

The fact that many high-quality LDPC codes have degree
distributions well fitted by power-law functions stimulates us to
test scale-free networks as error-correcting codes. Following the
approach of Newman {\it et al.}~\cite{New2}, we write the
generating function for a general scale-free network in the form:
\begin{equation}\label{eq2}
G(x) = \sum_{k=0}^{d_l-1}a_kx^k
+\sum_{k=d_l}^{d_{max}}p(k)x^k,
\end{equation}
where the fraction of nodes with degree $k \geq d_l$ follows a
power law $p(k)=Ck^{-\gamma}$, and the terms $a_k$ with low degree
$k < d_l$ are separated from the second sum to allow for possible
deviation from the power law for small $k$, as is often the case
for general scale-free networks. The generating function in
Eq.~(\ref{eq2}), however, contains too many parameters to be
amenable to numerical optimization unless $d_l$ is sufficiently
small. To reduce the number of parameters while still retaining
the possibility that the distribution for low degrees may not obey
an exact power law, we instead choose the following generating
function:
\begin{equation}\label{eq3}
G(x) = \sum_{k=0}^{d_{max}} p(k)x^k,
\end{equation}
where $p(k) = C(k+\alpha )^{-\gamma}$. If $\alpha <0$ ($\alpha
>0$), the degree distribution for small $k$ lies above (below) the
power-law function $k^{-\gamma}$. We henceforth use
Eq.~(\ref{eq3}) to generate a variable-node degree distribution of
our code and optimize the parameters $\alpha$ and $\gamma$ to
achieve the best performance.

Some empirical results known about LDPC codes help us to further
refine our code. The most well known findings related to features
of good LDPC codes may be that the variable nodes of degree one
should be removed since they do not contribute to error correction
and that the codes with almost uniform check-node degree yield
good performance \cite{Rich,Chung,Mac2}. Taking these into
account, we let the sum in Eq.~(\ref{eq3}) start from $k=2$, and
restrict the check-node degree to two consecutive integers: the
generating function for the check-node degree is written as
$F(x)=bx^i +(1-b)x^{i+1}$, where the parameters $b$ and $i$ are
easily determined once a variable-node degree distribution is
selected. This choice of the check-node degree distribution
enables us to design a code without restrictions on $d_{max}$ for
any given code rate; this property, however, is not shared by the
right-regular sequence \cite{Shok} for which $d_{max}$ is allowed
to have only a special set of values.

The performance of an LDPC code over a binary erasure channel can
be evaluated by the density evolution method \cite{Rich} as
follows. Let $\delta$ be the erasure probability of a given
channel, and consider a code with a degree distribution pair
$\lambda (x)=\sum \lambda _kx^{k-1}$ and $\rho (x)=\sum \rho
_kx^{k-1}$, where $\lambda _k$ ($\rho _k$) is the fraction of
edges connected to a variable (check) node of degree $k$. Note
that the distribution here is defined in terms of the fraction of
edges, not the fraction of nodes as before. Then, if the belief
propagation algorithm is used for decoding, the messages passed
between the variable and check nodes are known to evolve as
\cite{Rich,Luby}
\begin{equation}\label{eq4}
x_l=x_0\lambda (1-\rho (1-x_{l-1})),
\end{equation}
where $x_l$ denotes the expected fraction of erasure messages at
the $l$th iteration and $x_0$ is its initial value given by
$x_0=\delta$. The recovery of original data is successfully done
if $x_l$ converges to zero. The threshold $\delta ^*$, defined by
the supremum of all $\delta$ that result in successful decoding,
tells the code's performance.  For a given code rate $R$, the
threshold is upper bounded by $1-R$ \cite{Luby}.

\begin{figure}
\includegraphics[width=0.9 \columnwidth]{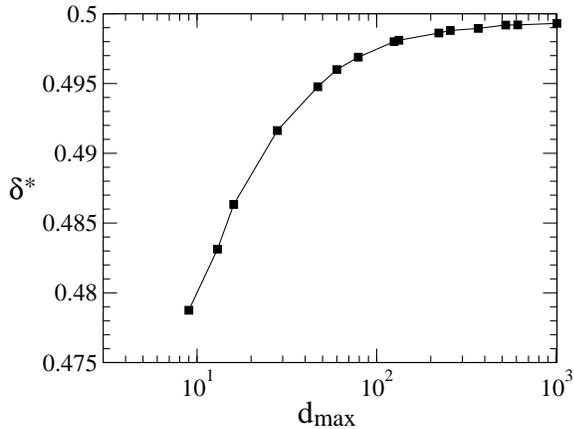}
\caption{\label{fig2} Error-correction capability of optimized
scale-free networks over a binary erasure channel. The code rate
is $R=0.5$. }
\end{figure}

With the help of the above density evolution method, we calculate
the error-correction capability of the scale-free networks given
in the form of Eq.~(\ref{eq3}). The results are shown in
Fig.~\ref{fig2} as a function of the maximum variable-node degree
$d_{max}$, where the code rate is fixed at $R=0.5$. It is seen
that the threshold erasure probability $\delta ^*$ increases as
the maximum variable-node degree increases. For large $d_{max}$,
the threshold almost reaches the theoretical upper bound $1-R$,
indicating that the error-correction capability of our code is
very good. For comparison, we have also studied the
error-correction capability of codes that have degree
distributions other than the power-law distribution, namely the
exponential distribution of the form $p(k)\sim e^{-\beta (k+\alpha
)}$ and the Gaussian distribution of the form $p(k)\sim e^{-\beta
(k+\alpha )^2}$. We find that the threshold for the exponential
distribution rapidly increases with $d_{max}$ and converges to
0.465, a value much lower than the threshold for the power-law
distribution. The case for the Gaussian distribution is observed
to exhibit a similar behavior with a similar, low convergence
limit.

\begin{table}[b] 
\caption{\label{tab1}Error-correction capabilities of scale-free
networks (SFN) and the Tornado code \cite{Shok} of rate $R=0.5$,
and parameters $\gamma$ and $\alpha$ optimizing the SFN.
The optimization was done by using a direction set method.}
\begin{ruledtabular}
\begin{tabular}{rcccccc}
\multicolumn{3}{c}{ } &\multicolumn{2}{c}{SFN}
&\multicolumn{2}{c}{Tornado \cite{Shok}} \\
\multicolumn{1}{c}{$d_{max}$} & $\gamma$ & $\alpha$ & $\delta ^*$
& $\langle k\rangle$
& $\delta ^*$ & $\langle k\rangle$ \\
\hline
9 & 1.347 & -1.473 & 0.47875 & 2.97 & 0.44546 & 3 \\
16 & 1.788 & -1.102 & 0.48633 & 3.30 & 0.46950 & 3.5 \\
28 & 2.024 & -0.868 & 0.49163 & 3.57 & 0.48235 & 4 \\
47 & 2.088 & -0.775 & 0.49477 & 3.88 & 0.48960 & 4.5 \\
79 & 2.084 & -0.753 & 0.49689 & 4.24 & 0.49380 & 5 \\
133 & 2.080 & -0.741 & 0.49810 & 4.60 & 0.49628 & 5.5 \\
222 & 2.086 & -0.712 & 0.49862 & 4.94 & 0.49776 & 6 \\
368 & 2.081 & -0.698 & 0.49895 & 5.31 & 0.49865 & 6.5 \\
610 & 2.076 & -0.691 & 0.49920 & 5.68 & 0.49918 & 7 \\
1009 & 2.073 & -0.687 & 0.49931 & 6.02 & 0.49951 & 7.5 \\
\end{tabular}
\end{ruledtabular}
\end{table}

To more clearly demonstrate the high performance of codes on
scale-free networks, we compare them with the Tornado code
\cite{Luby,Shok}. The threshold of our code is presented in Table
\ref{tab1} along with the parameters $\alpha$ and $\gamma$ that
maximize the code's performance. Table \ref{tab1} shows that our
code yields better performance than the Tornado code for $d_{max}$
smaller than about 1000. Also shown in Table \ref{tab1} is the
average variable-node degree $\langle k\rangle$ of the two codes.
From a practical viewpoint, it is important to design a code that
yields good performance for small $\langle k\rangle$, since the
physical complexity of a code, which grows with increasing
$\langle k\rangle$, limits the hardware implementation of the
code. For this reason, our code seems to be better suited to
applications than the Tornado code.

\begin{figure}
\includegraphics[width=0.95 \columnwidth]{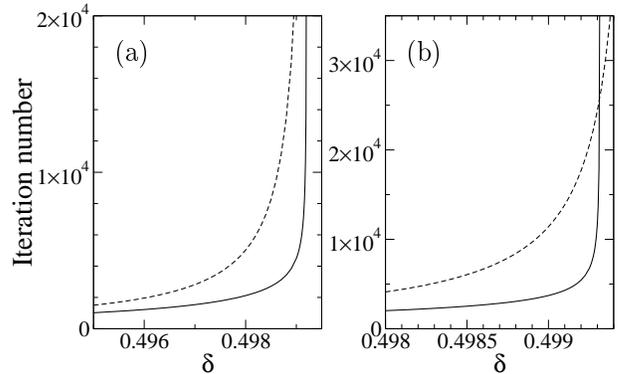}
\caption{\label{fig3}
Iteration numbers of optimized scale-free networks
(solid curve) and the Tornado code (dashed curve) for maximum
degrees (a) $d_{max}=610$ and (b) $d_{max}=1009$. The criterion
for convergence of decoding is set to be $x_l < 1\times 10^{-6}$.
}
\end{figure}

Another merit of our code is that the iteration number required
for convergence of decoding is very small. The iteration numbers
of our code and the Tornado code are compared in
Fig.~\ref{fig3}(a) as a function of the erasure probability for
the case of $d_{max}=610$, which clearly shows that our code has a
smaller iteration number than the Tornado code in the whole region
of $\delta$. Even for the case of $d_{max}>1000$ where the Tornado
code has a little higher threshold than our code, the iteration
number is smaller for our code than for the Tornado code over a
broad region of $\delta$, except near the threshold
[Fig.~\ref{fig3}(b)]. For an early convergence of decoding
processes, each node needs to gather messages from other nodes
quickly. This implies that graphs with smaller diameter may be
more advantageous to reducing the iteration number. This in turn
suggests that scale-free networks, which are known to have a very
small diameter $d \sim \ln{\ln{N}}$ \cite{Cohen} where $N$ is the
number of nodes, may require a smaller iteration number than
regular random networks or small-world networks \cite{Watt}.

The error-correction capability of codes on scale-free networks can be
further enhanced by adjusting the degree distribution, especially in the
low degree region, so that it more closely models realistic
scale-free networks whose degree distribution does not necessarily
follow a power law for small $k$. While doing this, we try to keep
as small as possible the number of parameters added to the generating
function. After a number of numerical simulations we have found the
following generating function adequate enough for this purpose:
\begin{equation}\label{eq5}
G(x) = C\left[ w_2 p(2)x^2 +w_3 p(3)x^3 +\sum_{k=4}^{d_{max}} p(k)x^k
\right] ,
\end{equation}
where $p(k) = (k+\alpha )^{-\gamma}$. The performance of this code
is displayed in Table \ref{tab2}, which shows that by adding two
new parameters, $w_2$ and $w_3$, which permit the two lowest
degrees to vary from the power law, the performance of the code is
increased. Addition of more parameters is expected to give rise to an
increased performance, but at the expense of rendering the
optimization process more time consuming.

\begin{table}
\caption{\label{tab2} Performance of codes on scale-free networks
[Eq.~(\ref{eq5})]. The same optimization parameters $\alpha$ and
$\gamma$ as in Table \ref{tab1} are used.}
\begin{ruledtabular}
\begin{tabular}{rcccc}
$d_{max}$ & $w_2$ & $w_3$ & $\delta ^*$ & $\langle k\rangle $ \\
\hline
222 & 1.004 & 0.983 & 0.49885 & 4.94 \\
368 & 1.004 & 0.982 & 0.49923 & 5.31 \\
610 & 1.005 & 0.982 & 0.49945 & 5.68 \\
1009 & 1.005 & 0.983 & 0.49955 & 6.01 \\
\end{tabular}
\end{ruledtabular}
\end{table}

In summary, we have found that many high-performance LDPC codes
possess degree distributions well fitted by power-law functions
with exponents close to 2. Based on this finding, we have
developed codes on scale-free networks that have very good
error-correction capabilities. The codes with power-law degree
distribution yield better performance than those with exponential
and Gaussian degree distributions that have fast decreasing tails.
It also would be interesting to study the effect of degree correlations
on the performance of a code, which is left as future work.
As good error-correcting codes, the codes on scale-free networks
could find lucrative applications in areas as diverse as wireless
communication, media and data transfer over the Internet, and
storage.

\end{document}